\documentclass[aps,prl,preprint]{revtex4}
\usepackage{graphicx}
\usepackage{bm}
\begin{document}
\title{Four-Wave Mixing With Self-Phase Matching due to Collective Atomic Recoil}
\author{G.R.M Robb \& B.W.J. M$^{\rm c}$Neil}
\affiliation{Department of Physics, John Anderson Building, \\
University of Strathclyde, Glasgow, G4 0NG, Scotland}
\date{\today}
\begin{abstract}
We describe a method for non-degenerate four-wave mixing in a cold sample of 4-level atoms. An integral part of the four-wave mixing process is a collective instability which spontaneously generates a periodic density modulation in the cold atomic sample with a period equal to half of the wavelength of the generated high-frequency optical field. Due to the generation of this density modulation, phase-matching between the pump and scattered fields is not a necessary initial condition for this wave-mixing process to occur, rather the density modulation acts to ``self phase-match'' the fields during the course of the wave-mixing process. We describe a one-dimensional model of this process, and suggest a proof-of-principle experiment which would involve pumping a sample of cold Cs atoms with three infra-red pump fields  to produce blue light.
\end{abstract}

\maketitle

Many methods of frequency conversion, and in particular frequency upconversion, can generally be divided into two broad categories. The first is the conventional nonlinear-optical method of frequency upconversion, which involves generating a high-frequency polarisation in an optical medium and is reliant on the internal structure of the atoms comprising the medium. The second involves scattering of light from relativistic electron beams as e.g. in the free electron laser, where the frequency upshift occurs due to the Doppler upshift of the scattered radiation, and coherence of the scattered field develops through the generation of a periodic density modulation of the individual scatterers (electrons) with a period comparable to the scattered field wavelength.
In this letter a new mechanism for the generation of short wavelength radiation due to four-wave mixing in a cold atomic gas is described, which incorporates features of both the classes described above. The four-wave mixing process arises due to a process of collective atomic recoil, during which the four-wave mixing is accompanied by the formation of an atomic density grating on the short wavelength scale, similar to the Collective Atomic Recoil Laser (CARL) \cite{CARL}, the atomic analogue of the free-electron laser. An essential difference between the original (degenerate) CARL process and the process described in this letter is that in CARL, the scattered field has approximately the same frequency as the pump. In the process described here, which can be thought of as non-degenerate CARL, the scattered field and pump fields can have very different frequencies.

\begin{figure}[h]
\includegraphics[height=5cm,clip=true]{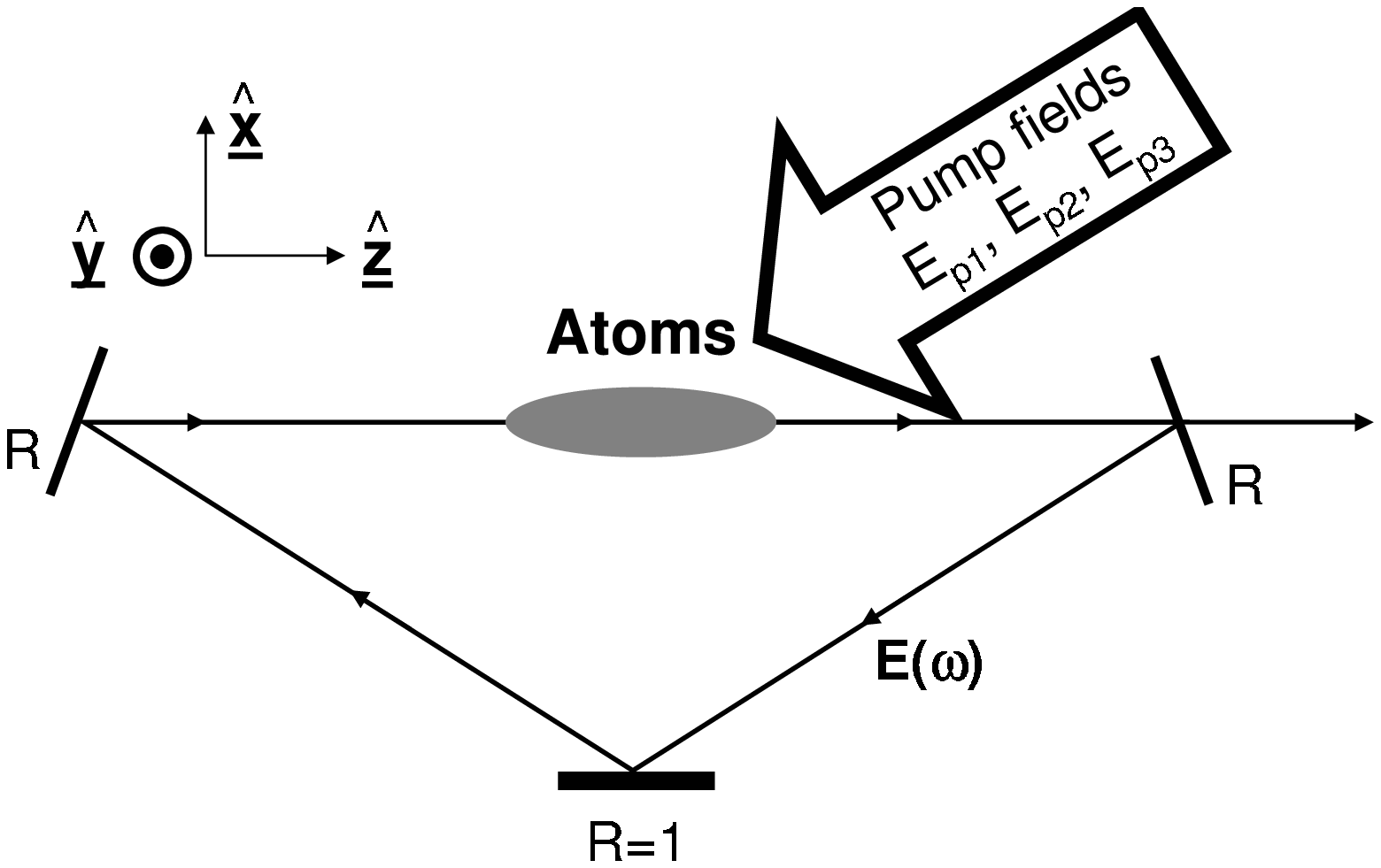}
\caption{Schematic diagram showing four-wave mixing in a high-finesse unidirectional cavity. The misalignment of the pump fields from the z-axis is exaggerated for clarity.}
\label{cavity}
\end{figure}

A one-dimensional model which describes the coupled evolution of a collection of 4-level atoms together with four optical fields is derived. The fields are assumed to be three strong pump fields with frequencies $\omega_{p1}$, $\omega_{p2}$ and $\omega_{p3}$ and a relatively weak probe field with frequency $\omega \approx \omega_{p1} + \omega_{p2} + \omega_{p3}$. For simplicity the optical fields and the atomic centre-of-mass motion are described classically. Only sum-frequency generation is discussed in this letter, although the same model can also be used to describe difference frequency generation.
For the purposes of this letter, in order to demonstrate the existence of the four-wave-mixing process, we consider the scenario depicted in fig.~\ref{cavity}, where the atomic sample is enclosed in a unidirectional ring cavity.
Although the use of the cavity is primarily a simplification of the model allowing the neglect of propagation effects, it is also relevant to many situations of experimental interest, there having been several recent experimental studies of cold atomic samples in optical cavities \cite{Kruse,Courteille,Vuletic,Hemmerich}. In cases where the finesse of the cavity is sufficiently high that the atoms interact effectively with only one mode of the cavity, it is possible to neglect the spatial evolution of the scattered field and consider only the temporal evolution of a single cavity mode amplitude. The configuration is assumed to be similar to that shown in figure~\ref{cavity}, with the short-wavelength cavity mode propagating almost opposite to the pump fields, which are assumed to be undepleted for simplicity. 

\begin{figure}[h]
\includegraphics[height=4cm,clip=true]{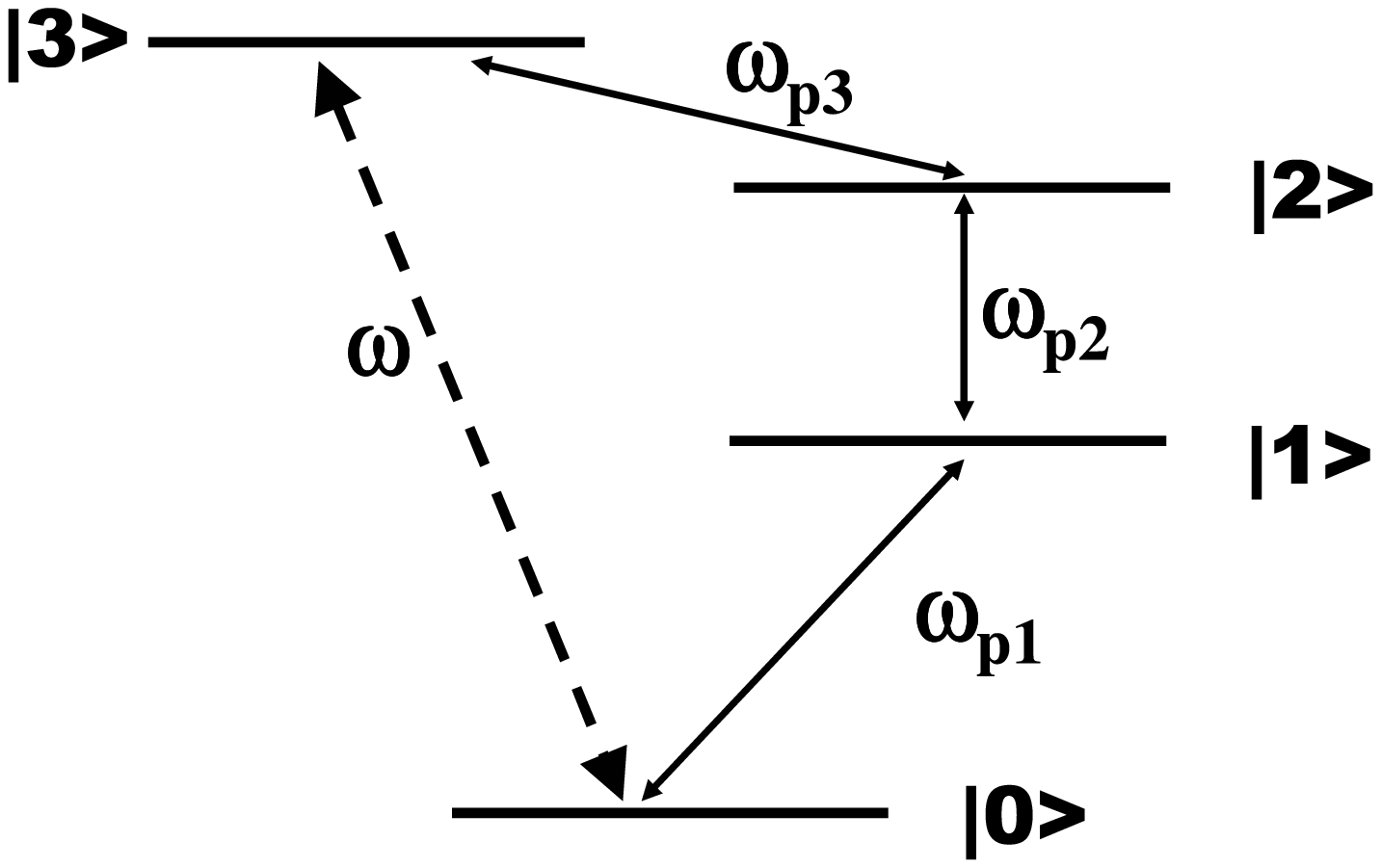}
\caption{Schematic of atomic energy level configuration}
\label{structure}
\end{figure}

The cold atomic sample is assumed to be composed of atoms with a four-level internal structure. The analysis which follows is not dependant on the specific form of this structure. The particular form we consider in this paper is shown in figure~\ref{structure}. The main reason for this choice is that it represents a good model for alkali atoms such as Cs and Na which are commonly used in cold atom experiments. Transitions  $\left| 1 \right \rangle \rightarrow \left| 0 \right \rangle$, $\left| 2 \right \rangle \rightarrow \left| 1 \right \rangle$, $\left| 3 \right \rangle \rightarrow \left| 2 \right \rangle$ and $\left| 3 \right \rangle \rightarrow \left| 0 \right \rangle$ are dipole coupled, and the transitions $\left| 2 \right \rangle \rightarrow \left| 0 \right \rangle$ and $\left| 3 \right \rangle \rightarrow \left| 1 \right \rangle$ are dipole forbidden. The high-frequency field generated by the four-wave mixing process will have a frequency $\omega \approx  \omega_{p1} + \omega_{p2} + \omega_{p3}$.

The axial ($z$) component of the force on each atom is given by
\begin{equation}
\label{force}
F_z = {\bm d} . \frac{\partial {\bm E}}{\partial z}
\end{equation}
where the dipole moment of each atom can be written on terms of the density matrix elements $\rho_{jk}$ as
\[
{\bm d} = ( {\bm \mu}_{10} \rho_{10} 
+ {\bm \mu}_{21} \rho_{21} + {\bm \mu}_{32} \rho_{32} + {\bm \mu}_{30} \rho_{30} + c.c. )
\]
where ${\bm \mu}_{jk} = {\bm \mu}_{kj}$ and $\rho_{jk} = \rho^*_{kj}$.
For simplicity it is assumed that the optical field ${\bm E}$ is linearly polarised in the direction $\hat{\bm y}$ and the direction of the dipole matrix elements ${\bm \mu}_{jk}$ are parallel to that of the electric field. Therefore
\begin{equation}
\label{defd}
{\bm d} = ( \mu_{10} \rho_{10} 
+ \mu_{21} \rho_{21} + \mu_{32} \rho_{32} + \mu_{30} \rho_{30} + c.c. ) \hat{\bm y}. 
\end{equation}
Assuming the density matrix elements and the total optical electric field can be written as
\begin{eqnarray}
\rho_{10} &=& s_{10} e^{-i (k_{p1} z + \omega_{p1} t)} \nonumber \\
\rho_{21} &=& s_{21} e^{-i (k_{p2} z + \omega_{p2} t)} \nonumber \\
\rho_{32} &=& s_{32} e^{-i (k_{p3} z + \omega_{p3} t)} \nonumber \\
\rho_{20} &=& s_{20} e^{-i \left[(k_{p1}+k_{p2}) z + (\omega_{p1} + \omega_{p2}) t\right]} \nonumber \\
\rho_{31} &=& s_{31} e^{-i \left[(k_{p2}+k_{p3}) z + (\omega_{p2} + \omega_{p3}) t\right]} \nonumber \\
\rho_{30} &=& s_{30} e^{-i \left[(k_{p1}+k_{p2}+k_{p3}) z + (\omega_{p1} + \omega_{p2} + \omega_{p3}) t\right]} \nonumber \\
E(z,t) &=& \left[  \left( A_{p1} e^{-i (k_{p1} z + \omega_{p1} t)} + A_{p2} e^{-i (k_{p2} z + \omega_{p2} t)} + A_{p3} e^{-i (k_{p3} z + \omega_{p3} t)} \right) + A(z,t) e^{i (k z - \omega t)} + c.c. \right] \label{defE}
\end{eqnarray}
where $k_{p1} = \omega_{p1}/c$, $k_{p2} = \omega_{p2}/c$, $k_{p3} = \omega_{p3}/c$, $k=\omega/c$ and $\omega \approx \omega_{p1}+\omega_{p2}+\omega_{p3}$, then after substituting for $\rho_{10}$, $\rho_{21}$, $\rho_{32}$ and $\rho_{30}$ in eq.(\ref{defd}), substituting ${\bm d}$ and ${\bm E}$ in eq.(\ref{force}) and neglecting fast-varying terms we eventually obtain
\begin{equation}
\label{force_temp}
F_z = i \hbar k \left( \Omega s_{30}^* e^{2 i k z} - c.c. \right)
\end{equation}
where $\Omega = \mu_{30} A / \hbar$ and we have assumed non-resonant excitation of the atomic sample by the pump beams, which are assumed to be much stronger than the high-frequency scattered field and are far-detuned from resonance. The assumption of non-resonant excitation means that we can assume negligible population in all but the ground state i.e. $\rho_{00} \approx 1, \rho_{11}=\rho_{22}=\rho_{33} \approx 0$, and adiabatically eliminate the coherences, so that for far-detuned pump fields, $s_{10} \approx \frac{\mu_{10} A_{p1}}{\hbar \Delta_{10}}$, $s_{21} \approx \frac{\mu_{21} A_{p2}}{\hbar \Delta_{21}}$ and $s_{32} \approx \frac{\mu_{32} A_{p3}}{\hbar \Delta_{32}}$, where $\Delta_{10}=\omega_{p1}-\omega_{10}$, $\Delta_{21}=\omega_{p2} - \omega_{21}$ and $\Delta_{32}=\omega_{p3}-\omega_{32}$. The possibility of enhancing the interaction via optimisation of the coherence $s_{30}$ will be the subject of a future study.

The evolution of the electromagnetic field is determined by Maxwell's wave equation
\[
\left( \nabla^2 - \frac{1}{c^2} \frac{\partial^2 }{\partial t^2} \right) {\bm E} = 
\frac{1}{\epsilon_0 c^2}
\frac{\partial^2 {\bm P}}{\partial t^2}
\]
where 
\[
{\bm P} = \sum_j^N {\bm d}_j  \delta (\bm{r} - {\bm r}_j(t)) .
\]
where ${\bm d}_j$ is the dipole moment of the $j$th atom, $j=1...N$, $N$ is the number of atoms, $\bm{r}$ is a position vector and $\bm{r}_j$ is the position of the $j$th atom. Assuming that the pump fields are undepleted, so that only the high-frequency field with frequency $\omega$ evolves, and that this field is resonant with a single mode of the cavity, then using the definitions of ${\bm E}$ and ${\bm d}_j$ in eq.(\ref{defE}) and eq.(\ref{defd}) respectively, assuming $\frac{\partial^2 {\bm P}}{\partial t^2} \approx - ( \omega)^2 \bm{P}$, applying the slowly varying envelope approximation (SVEA), averaging over an area $S$ in the $x-y$ plane and the cavity length ${\cal L}$, we obtain
\begin{equation}
\label{dadt_temp}
\frac{d  A(t)}{d t} = \frac{i \omega \mu_{30} n c}{2 \epsilon_0 c} \left \langle s_{30} e^{- 2 i k z} \right \rangle - \kappa A(t)
\end{equation}
where a damping term $-\kappa A$, has been added to represent mirror losses from the cavity, where $\kappa = \frac{c(1-R)}{\cal L}$ is the cavity linewidth and $R$ is the mirror reflectivity.

In the limit of non-resonant excitation, the coherence $s_{30}$ as a constant quantity $s_{30} = - \frac{\Omega_{p1} \Omega_{p2} \Omega_{p3}}{\Delta_{10} \Delta_{20} \Delta_{30}}$, where $\Delta_{20}=\Delta_{10}+\Delta_{21}$ and $\Delta_{30}=\Delta_{20}+\Delta_{32}$. The equations describing the system therefore reduce to
\begin{eqnarray}
\frac{d^2 \theta_j(t)}{d t^2} &=& - \frac{\omega^2 \mu s_{30}}{m c^2} \left( A e^{i \theta_j} + c.c. \right) \label{d2theta_dt2}\\
\frac{d A(t)}{d t} &=& \frac{\mu c \omega n s_{30}}{2 \epsilon_0 c} \left \langle e^{- i \theta } \right \rangle 
- \kappa A(t)
\end{eqnarray}
where $\theta = 2 k z$.

\begin{figure}[h]
\includegraphics[width=8cm,clip=true]{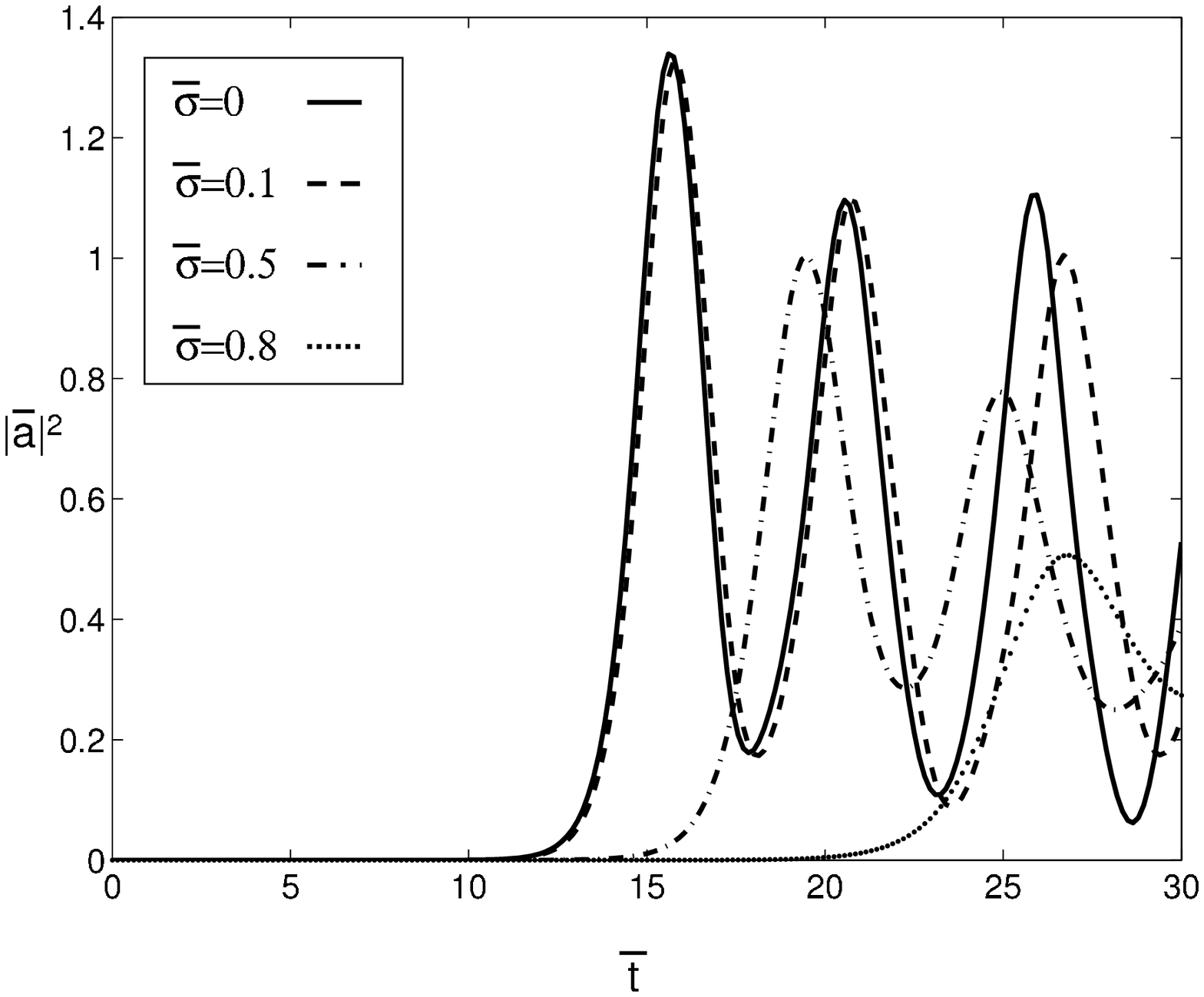}
\caption{Scaled intracavity intensity ($|\bar{a}|^2$) as a function of $\bar{t}$ for values of $\bar{\sigma}$.}
\label{intvst}
\end{figure}

Using the scaling 
\[
\bar{p} = \frac{p}{\hbar k \rho} \;\;\;,\;\;\; \bar{t} = \omega_r \rho t \;\;\;,\;\;\; \bar{z} = \omega_r \rho z / c,
\]
\[
\bar{a} = -i \sqrt{\frac{2 \epsilon_0}{n \hbar \omega \rho}} A \;\;\;,\;\;\; 
\rho = \left ( \frac{\mu^2 \omega n s_{30}^2}{2 \epsilon_0 \hbar \omega_r^2} \right)^{1/3}, \bar{\kappa} = \frac{\kappa}{\omega_r \rho}
\]
where $ \omega_r = \frac{ 2 \hbar k^2}{m}$, $m$ is the atomic mass, $n = \frac{N}{S {\cal L}} = n_s \frac{L}{\cal L}$ is the atomic sample density with respect to the cavity volume and $n_s$ is the true number density of the atomic sample, then the scaled equations become those of the high-gain free electron laser (FEL) \cite{FEL} (when the cavity damping term $\kappa \rightarrow 0$) and the (degenerate) CARL \cite{CARL} i.e.
\begin{eqnarray}
\frac{d \theta_j}{d \bar{t}} &=& \bar{p}_j \label{mf1} \\
\frac{d \bar{p}_j}{d \bar{t}} &=& - \left( \bar{a} e^{i \theta_j} + c.c. \right) \\
\frac{d \bar{a}}{d \bar{t}} &=& \left \langle e^{-i \theta} \right \rangle - \bar{\kappa} \bar{a} \label{mf3}.
\end{eqnarray} 
The initial temperature, T, of the atoms is represented using a Gaussian distribution of atomic momenta with half-width $\bar{\sigma}=\sqrt{\frac{2 k_B T}{m}}$, such that
\[
\left \langle f(\theta,\bar{p}) \right \rangle  = \frac{1}{2 \pi} \frac{1}{\sqrt{2 \pi} \bar{\sigma}} \int_{0}^{2 \pi} \int_{-\infty}^{\infty} f(\theta, \bar{p}) e^{-\frac{\bar{p}^2}{2 \bar{\sigma}^2}}  d \theta \; d \bar{p}
\]
and we assume that the finesse of the cavity is sufficiently high that we are operating in the ``good cavity limit'', where $\kappa \ll 1$ and the effect of cavity losses on the interaction dynamics is negligible. A numerical calculation of the scaled intracavity mode intensity, $|\bar{a}|^2$ is shown in fig.~\ref{intvst} as a function of $\bar{t}$ for different values of scaled thermal velocity spread, $\bar{\sigma}$. The initial conditions used were $|a(\bar{t}=0)|=1 \times 10^{-5}$ and uniformly spaced atoms such that $\left \langle e^{-i \theta} \right \rangle = 0$. It can be seen that maximum growth of the cavity mode (i.e. the high-frequency scattered field) occurs when $\bar{\sigma} \rightarrow 0$ i.e. for a perfectly cold beam with zero temperature, as would be expected.  It can also be seen from fig.~\ref{intvst} that the growth rate of the instability decreases with increasing $\bar{\sigma}$.
Consequently in the good cavity limit of negligible cavity losses, a system of initially unbunched atoms illuminated by the pump beams is unstable, leading to exponential amplification of both the short wavelength field and an atomic density modulation with a spatial period $\lambda/2$, where $\lambda = 2 \pi / (k_{p1}+k_{p2}+k_{p3})$, demonstrating the possibility of amplifying a short wavelength optical field via a four-wave mixing process which self-phase matches due to collective atomic recoil. For atomic temperatures sufficiently low that $\bar{\sigma} \leq 0.1$, the effects of atomic temperature on the collective instability are negligible. 

As an illustrative example we consider the case of a cylindrical sample of $N \approx 10^6$ Cs atoms illuminated by three infra-red pump lasers being used to produce visible blue light as illustrated in fig.~\ref{cavity}. The atomic energy levels $|0  \rangle$, $|1  \rangle$, $|2 \rangle$ and $|3 \rangle$ correspond to the $6S_{1/2}$, $6P_{3/2}$, $7S_{1/2}$ and $7P_{3/2}$ levels of the Cs atom respectively. The Einstein coefficients of transitions $|1 \rangle \rightarrow |0 \rangle$, $|2 \rangle \rightarrow |1 \rangle$, $|3 \rangle \rightarrow |2 \rangle$ and $|3 \rangle \rightarrow |0 \rangle$ are $A_{10}=3.3 \times 10^7 \mbox{s}^{-1}$, $A_{21}=1.2 \times 10^7 \mbox{s}^{-1}$, $A_{32}=4.0 \times 10^6 \mbox{s}^{-1}$ and $A_{30}=4.2 \times 10^6 \mbox{s}^{-1}$ respectively \cite{Kurucz}. The sample is assumed to have a length $L \approx 200 \mbox{$\mu$ m}$ and radius $R \approx 40 \mbox{$\mu$ m}$, giving an atomic sample density $n_s \approx 1 \times 10^{18} \mbox{m}^{-3}$. The cavity is assumed to have length ${\cal L} =10 \mbox{cm}$ and mirror transmittivity $T_c = 3 \times 10^{-5}$, so the atomic density relative to the cavity is consequently $n = n_s L / {\cal L} = 2 \times 10^{15} \mbox{m}^{-3}$. The three pump fields have wavelengths $\lambda_{p1} \approx 852 \mbox{nm}$, $\lambda_{p2} \approx 1.47 \mbox{$\mu$ m}$, $\lambda_{p3} \approx 2.93 \mbox{$\mu$ m}$ and the scattered field will have a wavelength $\lambda \approx 455 \mbox{nm}$. The pump fields are assumed to be sufficiently far-detuned that the approximation of non-resonant excitation and negligible population in all levels apart from the ground state is valid. For this example we have chosen $\Delta_{10} = 5000 A_{10}$, $\Delta_{21}=25 \Delta_{10}$ and $\Delta_{32}=-25 \Delta_{10}$. Consequently $\Delta_{20} \approx \Delta_{21}$ and $\Delta_{30} = \Delta_{10}$. The Rabi frequencies have been chosen for simplicity as $\Omega_{p1} = \Delta_{10}/5$. $\Omega_{p2} = \Delta_{21}/5$ and $\Omega_{p3} = |\Delta_{32}|/5$, which corresponds to pump intensities of $I_{p1}=8.9 \times 10^3 \mbox{W cm}^{-2}$, $I_{p2}=4.3 \times 10^6 \mbox{W cm}^{-2}$ and $I_{p3}=1.6 \times 10^6 \mbox{W cm}^{-2}$ respectively. Using these parameters, the parameter $\rho \approx 47$ and $\omega_r = 1.8 \times 10^{5} \mbox{rad s$^{-1}$}$. The recoil temperature of the atoms is $\approx 1.4 \mbox{$\mu$K}$. The classical treatment of the atomic dynamics in this letter is strictly valid only for temperatures above the recoil temperature. Assuming $T \approx 7 \mbox{$\mu$K}$, the corresponding value of the scaled thermal momentum spread is $\bar{\sigma} \approx 0.1$. From fig.~\ref{intvst}, it can be seen that for this value of $\bar{\sigma}$ the behaviour of the system is very close to that in the cold beam limit, and the effects of thermal spread are almost negligible. For this example the short-wavelength scattered radiation in the cavity is amplified to an intensity of $1.3 \times 10^3 \mbox{W cm}^{-2}$ at saturation ($39 \mbox{mW cm}^{-2}$ transmitted from the cavity) after a time of $\approx 1.8 \mu\mbox{s}$. 

\begin{figure}[h]
\includegraphics[height=4cm,clip=true]{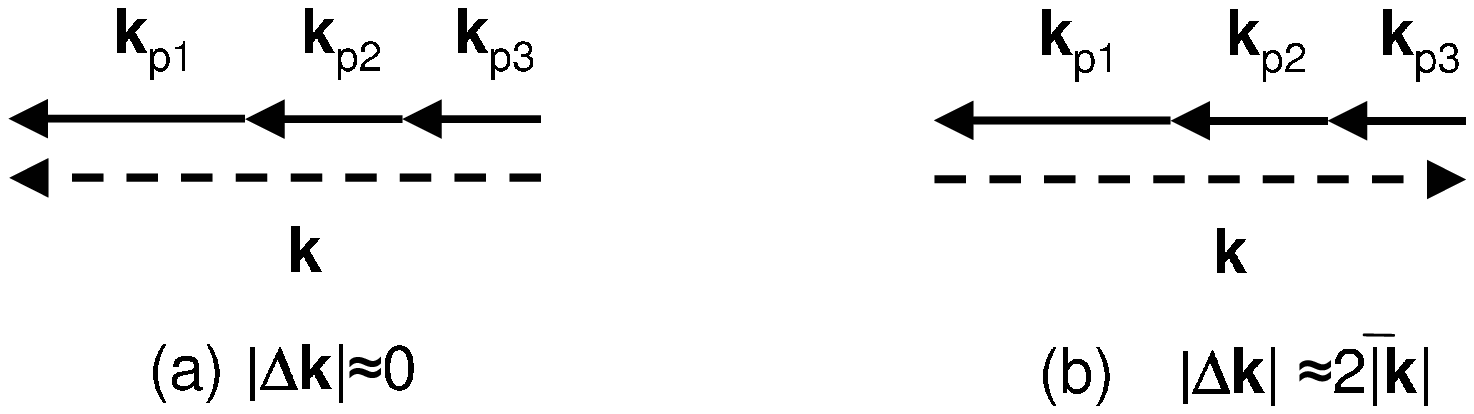}
\caption{Schematic showing configuration for (a) conventional four-wave mixing (b) four-wave mixing with self-phase matching}
\label{mismatch}
\end{figure}

In conventional four-wave mixing \cite{conventionalFWM}, the pump and scattered/generated fields are said to be phase-matched i.e. the wave vectors of the pump fields and the harmonic field satisfy ${\bm \Delta k} = {\bm k}_{p1}+ {\bm k}_{p2} + {\bm k}_{p3} - {\bm k} \approx 0$ (see fig.~\ref{mismatch}(a)), such that the coherence length $L_c = 2 \pi / | {\bm \Delta k}|$ is much larger than the generated radiation wavelength, $\lambda$. 
For situations where wavevector mismatch between the pump and scattered fields is large e.g. as in fig.~\ref{cavity} and fig.~\ref{mismatch}(b), the coherence length will be less than the radiation wavelength, $L_c < \lambda$, so the amount of high-frequency radiation generated by conventional four-wave mixing will be negligibly small as it varies as $\propto \left(L_c / L \right)^2$ \cite{conventionalFWM}, where $L$ is the length of the medium which is much greater than the coherence length, $L_c$. In the case of four-wave mixing due to recoil however, this is not the case because the degree of phase matching is proportional to the degree of atomic bunching. Although initially this is very small, as the collective instability develops the atoms form a strong density modulation with a spatial period equal to half of the wavelength of the high-frequency field, which reduces the phase mismatch. The effect of the collective instability due to recoil is therefore to ``self-phase match'' the pump fields and the scattered high-frequency field. Note that a wavevector mismatch, usually undesirable during a conventional four-wave mixing process, is necessary for self-phase matching via collective atomic recoil, because it is the physical origin of the force which moves the atoms and initiates the formation of the atomic density modulation.

In conclusion, it has been shown that a collection of cold four-level atoms illuminated by three pump laser fields can generate a fourth, short wavelength field via a four-wave mixing process. The novel feature of the nondegenerate four-wave mixing process considered here is that it does not require phase matching, but instead relies upon a collective instability, in which the atoms spontaneously form a periodic density modulation which acts to ``self-phase match'' the pump and scattered fields. A possible scheme for a proof of principle experiment involving a high-finesse cavity to contain a sample of cold Cs atoms and the scattered, short wavelength field is described. It is important to note that the phenomenon described in this paper is not critically dependant on the presence of the optical cavity. The cavity acts to enhance the atom-field interaction and therefore to relax the restrictions on some parameters e.g. the atomic temperature necessary in order to successfully demonstrate the effect in an experiment. It is anticipated that a superradiant or superfluorescent analogue of the phenomenon described here could occur in a sample of cold atoms in free space, where the propagation direction of the scattered field is determined by the major axis of the cold atomic sample. However, based on previous studies of the degenerate CARL process \cite{CARL,CARL-MIT}, it is anticipated that sub-recoil temperatures would be necessary for this to occur, and a consistent study would require a quantum-mechanical description of the atomic motion. It should also be noted that we have performed a number of simplifying assumptions e.g. non-resonant excitation, undepleted pump fields, in order to highlight the novel features of the physical processes described here i.e. non-degenerate wave-mixing with phase matching due to a collective instability, but it is certainly possible that these may not be the optimum conditions under which to observe the phenomena described here. Examples of extensions to the analysis presented in this letter are (i) resonant enhancement of the coherence $s_{30}$, which would seem to offer a way of enhancing the collective instability through increasing the coupling between the atoms and the scattered field (the coupling parameter $\rho \propto s_{30}^{2/3}$) and reducing the pump laser intensities required to observe the instability  (ii) the addition of atomic cooling forces as in \cite{Kruse,Courteille}, which tends to relax the restriction on atomic temperature needed to observe the collective instability and damps out the oscillatory behaviour of the interaction after saturation (see fig.~\ref{intvst}). Such extended analyses will be necessary to deduce the conditions for optimum generation of the short-wavelength field. This is an important issue, as phase matching is the main limiting factor to efficient generation of very short-wavelength radiation e.g. XUV \& X-ray radiation, so the method of frequency conversion described in this letter could form the basis of new methods for realising such sources. Another possible application of the four-wave mixing process described here could be to produce short period density gratings/modulations in atomic media using long-wavelength light (in the example given here, three infrared lasers produce a density grating with a spatial period of 228nm). 

\acknowledgments
The authors would like to thank N. Piovella, R. Bonifacio \& W.J. Firth for helpful discussions. GRMR acknowledges support from INTAS grant 211-855.

\end{document}